\documentclass[opre]{informs3nothing}  

\DoubleSpacedXI 

\usepackage{endnotes}
\usepackage{float}
\usepackage{hyperref}
\let\footnote=\endnote
\let\enotesize=\normalsize
\def\notesname{Endnotes}%
\def\makeenmark{$^{\theenmark}$}
\def\enoteformat{\rightskip0pt\leftskip0pt\parindent=1.75em
  \leavevmode\llap{\theenmark.\enskip}}

\usepackage{natbib}
\usepackage{booktabs}
\usepackage{multirow}

 \bibpunct[, ]{(}{)}{,}{a}{}{,}%
 \def\bibfont{\small}%

\newcolumntype{x}[1]{>{\hfil$\displaystyle} p{#1} <{$\hfil}} 

\TheoremsNumberedThrough     
\ECRepeatTheorems

\EquationsNumberedThrough    

\begin{document}


\RUNAUTHOR{Qureshi and Zaman}

\RUNTITLE{The Impact of COVID-19 on Sports Betting Markets}

\TITLE{The Impact of COVID-19 on Sports Betting Markets}

\ARTICLEAUTHORS{%
\AUTHOR{Khizar Qureshi}
\AFF{Massachusetts Institute of Technology, Cambridge, MA 02139, \EMAIL{kqureshi@mit.edu}} 
\AUTHOR{Tauhid Zaman}
\AFF{Yale University, New Haven, CT 06511, \EMAIL{tauhid.zaman@yale.edu}}
} 

\ABSTRACT{%
We investigate the impact of the COVID-19 pandemic on the betting markets of  professional and college sports. We find that during the pandemic, the moneyline betting markets of the National Basketball Association (NBA) became very inefficient.  During this period, if one bet uniformly on underdog teams in the NBA, one could achieve a 16.7\% profit margin. It is hypothesized that this inefficiency is due to the absence of live audiences during the NBA games. Such inefficiencies are not seen for any other sport.  Much of the inefficiency comes from a small fraction of games with moneyline odds in the 233 to 400 range.    We find that with clever strategies, one is able to achieve a 26-fold gain on an initial investment by betting on NBA underdogs during this time period.
}%


\KEYWORDS{probability, statistics, sports, COVID-19, moneyline bets} \HISTORY{}

\maketitle

%


\section{Introduction}
Over the past three decades, there has been rapid growth in the size of betting markets for professional and college sports. A survey by the American Gaming Association found that between 2018 and 2020 the legal U.S. sports betting market surged from \$6.6 billion to  \$25.5 billion  \citep{betting_growth}. Alongside the launch of several new sports betting markets, there was also adoption on a state level, most recently in Illinois and Colorado. Despite this increase in the popularity of sports betting, the general consensus has been that betting markets have stayed efficient.   There have been some studies finding specific inefficiencies embedded within a small set of  features or match-ups \citep{berkowitz,borghesi,gandar}, but nothing systematically inefficient. It is unlikely that these markets are profitable for amateur betters, unless there was some extreme exogenous event that disrupted either these markets or the sports themselves. 

The COVID-19 pandemic was just such an extreme event that affected nearly every professional and college sports league in the world in 2020 and 2021.  The National Basketball Association (NBA) and National Hockey League (NHL) had to suspend their on-going seasons when the pandemic began \citep{nba,nhl}.  Later, in the summer of 2020, the NBA resumed its season in an isolated environment where players were quarantined and no fans were allowed \citep{bubble}.  This isolation \emph{bubble} was the NBA's attempt to let the season continue safely.  While Major League Baseball (MLB) and the NHL resumed their 2020 seasons without fans, the National Football League (NFL) resumed with 13 teams allowing fans at partial capacity.  The NBA's 2021 season began in complete isolation and then gradually fans were allowed to attend the games, depending on the hometown city's policies.  COVID-19 affected other aspects of the sports besides fan attendance.  Game  schedules were altered to minimize the amount of travel for the teams.  Players who tested positive for COVID-19 (or had contact exposure), were excluded from participating in games. This often resulted in games where superstars or key players were missing.  

An interesting element of sports is their respective betting markets.  One popular type of bet in these markets is known as the \emph{moneyline}.  In this bet, the odds makers give payouts for each team in a game.  The team with the higher payout is known as the \emph{underdog} because they have a lower implied probability of winning, assuming the payouts are chosen to make the bet have non-positive mean payout.  In efficient markets, one cannot make a consistent profit by betting solely on underdogs or favorites.  An interesting question is what impact COVID-19 had on the efficiency of betting markets.  If COVID-19 created inefficiencies, in what sport did this occur, what was the nature of the inefficiencies,  and how much could one earn by betting in these inefficient markets?

In this work we analyze the impact of COVID-19 on the efficiency of  moneyline betting markets for a variety of sports.  Our main finding is that the NBA experienced incredible inefficiencies in its betting markets during COVID-19, while other major sports had markets that stayed efficient.  The inefficiency in the NBA market was so stark that simple betting strategies were profitable.  For instance, consider the strategy where each day one bets an equal amount on the underdog in each game.  Using this simple underdog strategy from when the NBA COVID-19 bubble began until the end of the 2020 season gave a return of 16.7\%.  More clever strategies that reinvest the winnings gave returns as high as 2,666\%.  We conduct a deeper analysis of the data to explain the nature of the NBA inefficiencies.  We also test different betting strategies to see which ones are most profitable in this market.

This paper is organized as follows.  We begin by reviewing related literature in Section \ref{sec:review}.  Then in Section \ref{sec:data} we describe our dataset on moneyline odds for various sports.  In Section \ref{sec:covid} we present a detailed analysis of the efficiency of the moneyline betting markets for all sports across multiple seasons, including during the COVID-19 pandemic.  We study different betting strategies given the moneyline inefficiencies in Section \ref{sec:betting}.  Finally, we conclude in Section \ref{sec:conclusion}.

\section{Related Literature}\label{sec:review}
Many studies have looked at the efficiency  in the sports betting markets. \cite{gandar} find that the home-field advantage is efficient for the NFL, MLB, and NBA in both the regular season and playoff games.  However, \cite{borghesi} finds that in the NFL the realized winning probability for the home-underdog is significantly larger during the later part of a season. \cite{woodland} found for college basketball and football that there is clear evidence of the favorite long shot bias in money-line betting markets and that betting on heavy favorites offers a near zero return over several years, suggesting that the markets for these two sports are efficient within transaction costs.  \cite{gray} find that in select seasons a few linear models can be used to generate significant profits in NFL betting markets, but for most years the market is efficient.

Much of the COVID-19 related sports research focused not on betting markets, but on the potential impacts of the virus on athletes at both the microscopic and macroscopic level. For example, \cite{verwoert} consider a tree-based protocol dictating the stratification of athletes from a cardiovascular perspective. The logic branches on asymptomatic/regional/local symptoms and hospitalization. Similarly, researchers have provided a cardio-vascular based risk-mitigation approach for the eligibility of an athlete to return to a sport depending on his or her state (positive test or waiting for test) \citep{schellhorn,baggish}. Meanwhile, \cite{wong} studied the transmission risk of COVID-19 in soccer based on the degree of contact between players and their actions. For example, using forwards and mid-fielders,  they detail the contagion risk for a player over various time intervals of the game.

\section{Moneyline Definitions and Datasets}\label{sec:data}

\subsection{Moneyline Bets}
We begin by presenting a brief description of moneyline bets.  In a  game between two teams, oddsmakers will set moneyline odds for each team.   If one team's odds are strictly greater than the other team's odds, we refer to the team with the larger odds as the \emph{underdog}  and the other team as the \emph{favorite}.  If both teams' odds are equal, we refer to them both as underdogs.  The moneyline bet works as follows.  Consider a game where team $u$  has moneyline odds $o_u$.  In typical moneyline bets we either have $o_u\leq-100$ or $o_u\geq 100$, and each case has a different payout.   First, we consider $o_u\geq 100$.   In this case, if one bets 100 USD on $u$ and the team wins, then one receives a payout of $100+o_u$ USD (a net profit of $o_u$ USD).  If on the other hand, $u$ loses, then one incurs a loss of 100 USD.  Second, we consider $o_u\leq -100$.   In this case, if one bets $o_u$ USD on $u$ and the team wins, then one receives a payout of $100+o_u$ USD (a net profit of $100$ USD).  If $u$ loses, then one incurs a loss of $o_u$ USD.  

 The implied probability of a team winning is determined by the moneyline odds and the payout of the moneyline bet.  Let us define the implied probability of winning for a team $u$ as $p_u$.  We can calculate the implied probability by setting the expected profit of a moneyline bet on the team equal to zero.  For $o_u\geq 100$, this gives $o_up_u -100(1-p_u)=0$ and for $o_u\leq -100$, this gives $100p_u +o_u(1-p_u)=0$.  Solving for $p_u$ gives
\begin{align}
p_u &= \frac{100}{100+|o_u|}, ~~~o_u\geq 100.\label{eqn:pdog}\\
p_u &= \frac{|o_u|}{100+|o_u|} ~~~o_u\leq -100.\label{eqn:pfav}
\end{align}

\subsection{Underdog Profit Margin}
We now consider  measures to quantify the efficiency of betting markets.  There are multiple ways to do this, but in this work we focus on one simple measure which will prove useful.  We assume a bettor places a 1 USD bet on the underdog in each game in a set of games.  Then the efficiency measure, which we refer to as the \emph{average underdog profit margin}, is the average profit margin per game using this underdog betting strategy.  To provide an expression for this measure, we define some terms.  For a team $u$, let $p_u$ be the implied win probability based on the moneyline odds, and let $q_u$ be the actual win probability.    Let $W_u$ be a Bernoulli random variable that is one if the team wins the game.  With our notation we have $\mathbf E[W_u] = q_u$.  Assume we bet 1 USD on the underdog in $n$ different games.  Then a simple calculation shows that the average underdog profit margin, which we define as $\alpha$, is equal to 
\begin{align}
\alpha = -1 + \frac{1}{n}\sum_{u=1}^n \frac{W_u}{p_u}\label{eq:upm}.
\end{align}
To gain insights into this measure, we take the expectation over the uncertainty in game outcome to obtain
\[
\mathbf E[\alpha] = -1 + \frac{1}{n}\sum_{u=1}^n \frac{q_u}{p_u}.
\]
We see from this expression that $\alpha$ measures how much the mean of the ratio $q_u/p_u$ deviates from one.  If the oddsmakers set the odds accurately so $p_u=q_u$, then $\alpha$ is zero.  If the oddsmakers are efficiently setting underdog moneyline odds, $\alpha$ should be negative, implying one cannot make a consistent profit by betting on underdogs. However, if there are many upsets and underdogs win more frequently than implied by the odds, $\alpha$ will be positive. 
\subsection{Data}
We collected the moneyline odds for thousands of games from a website that archived  this data for multiple sports across multiple seasons \citep{sportsbookreviewsonline}.  Our dataset covers professional sports leagues such as the NFL, NBA, NHL, and MLB, along with college sports such as NCAA Football (NCAAF) and NCAA Basketball (NCAAB).  For each game, we have the initial moneyline odds for both teams, the date of the game, and the game outcome.   In total we have moneyline data for over 130 thousand games spanning 14 years.  The complete dataset can be obtained from our project repository \citep{github}.

Our data covers seasons as far back as 2007, but the data for MLB begins at 2010.  For this reason, in our analysis we consider seasons between 2010 and 2021 for all sports.  Also, in order to exclude potentially erroneous records in our dataset, we require the following conditions for a game to be included:
\begin{enumerate}
    \item The favorite odds must be less than or equal to -100
    \item The underdog odds must be either greater than 100 or between -200 and -100.
\end{enumerate}
There are 109,249 games which satisfy these constraints in our data set and 40 games which do not.  Table \ref{table:data} contains a summary of the dataset. 

\begin{table}
\caption{The number of games for each sport in our moneyline odds dataset, along with the dates covered by each sport.}\label{table:data}

\centering
\begin{tabular}{@{}|l|l|l|c|@{}}
\toprule
\textbf{}      & \textbf{Start Date} & \textbf{End Date} & \textbf{Number of Games} \\ \midrule
\textbf{MLB}   & 2010-04-04          & 2020-10-27        & 25,599          \\\hline
\textbf{NBA}   & 2007-10-30          & 2021-03-04        & 17,196          \\\hline
\textbf{NCAAB} & 2007-11-05          & 2021-02-16        & 56,052          \\\hline
\textbf{NCAAF} & 2007-08-30          & 2020-09-26        & 10,672          \\\hline
\textbf{NFL}   & 2007-09-06          & 2021-02-07        & 3,740           \\\hline
\textbf{NHL}   & 2007-09-29          & 2021-03-22        & 17,080          \\ \bottomrule
\end{tabular}
\end{table}


\section{Statistical Analysis of Betting Market Inefficiencies During COVID-19}\label{sec:covid}
 COVID-19 disrupted certain games in our dataset, both in terms of game outcomes and betting market efficiency.  In this section we conduct a statistical analysis comparing these properties with respect to underdogs for COVID-19 games and normal season games.  To begin this analysis we had to determine which games to designate as COVID-19 games.  The NBA and NHL experienced mid-season pauses due to COVID-19.  The COVID-19 games for these sports were played after the resumption of the season.   For the NFL and NCAAF, the season was not paused, so their COVID-19 games started in the fall of 2020.   The NCAAB only cancelled the March Madness Tournament, but all regular season games were played before COVID-19.  Table \ref{table:pause} contains the start dates of the COVID-19 games and the number of COVID-19 games played for each sport, along with the number of games in a normal season for reference.  There is a range of games due to some teams receiving different treatment based on their win-loss records when COVID-19 began impacting game schedules.
\begin{table}[h]
\label{table:pause}
\begin{tabular}{|l|l|l|l|}
\hline
\textbf{Sport} & \textbf{COVID-19 Games } & \textbf{Number of Normal }  & \textbf{Number of COVID-19} \\ 
               &  \textbf{Start Date }    & \textbf{ Season Games}   & \textbf{Games}  \\ \hline
\textbf{NBA}   & 2020-07-30 & 81  & 64-72 \\ \hline
\textbf{NFL}   & 2020-09-10 & 16  & 17 \\ \hline
\textbf{NHL}   & 2020-08-01 & 68  & 57-63 \\ \hline
\textbf{MLB}   & 2020-07-23 & 162 & 60 \\ \hline
\textbf{NCAAF} & 2020-09-03 & 10-13  & 4-12 \\ \hline
\textbf{NCAAB} & N/A        & 25-35  & 25-35 \\ \hline
\end{tabular}
\caption{Start date of COVID-19 games, number of games per season, and number of COVID-19  games played for each sport in our dataset.}
\end{table}


\subsection{Underdog Profit Margin and Win Probability }
\begin{figure}
\begin{center}
\includegraphics[width=\textwidth]{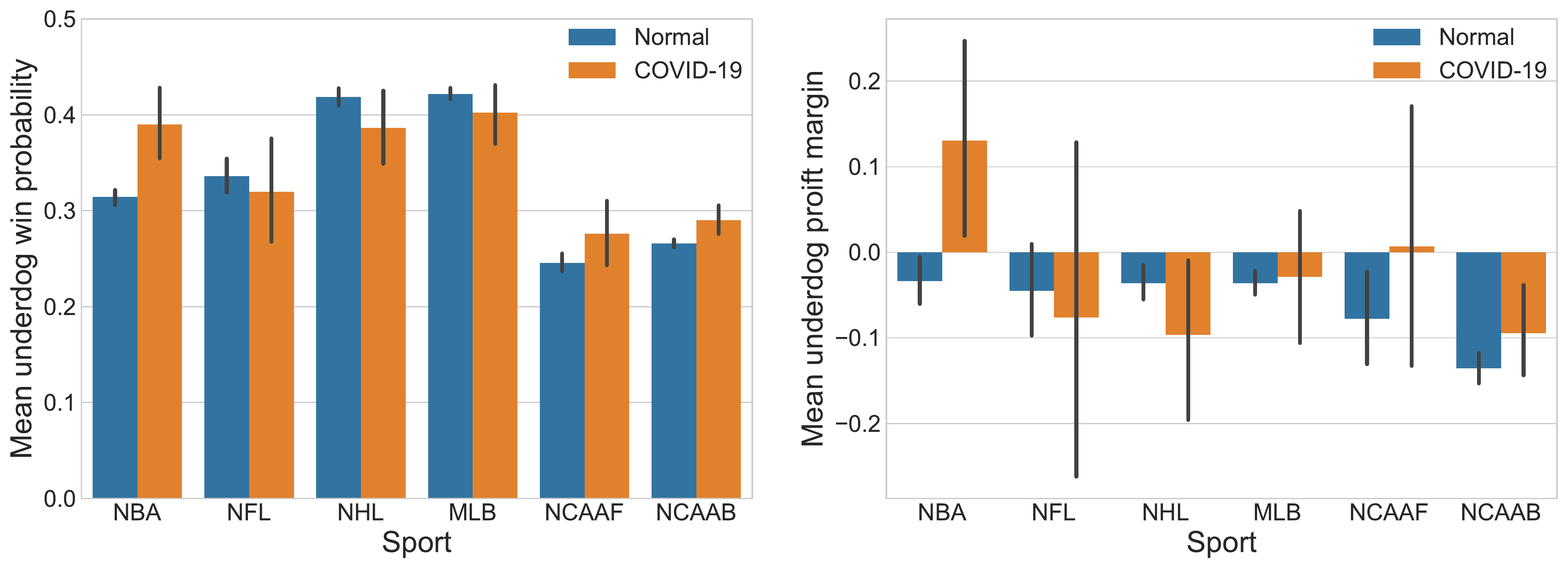}
\caption{Plot of (left) mean underdog win probability and (right) mean underdog profit margin grouped by sport and COVID-19 status.  The error bars represent 95\% confidence intervals.}
\label{fig:covid_sport}
\end{center}
\end{figure}

For each season we study two properties of underdogs:  their actual win probability, and the efficiency of their moneyline odds.  The underdog win probability shows how frequently upsets occurred.  The underdog profit margin provides a monetary value to these upsets.  Figure \ref{fig:covid_sport} shows the average underdog win probability and profit margin for the sports during normal and COVID-19 games.   It can be seen that the NBA shows the largest increase in win probability and profit margin for underdogs during COVID-19.  The NBA underdog win probability goes from approximately 0.3 to 0.4.   For all other sports, there is not such a visible difference.  In fact, if we look at the underdog profit margin in Figure \ref{fig:covid_sport}, we see that the NBA is the only sport showing a substantial positive mean during COVID-19.  

We tested the significance of the differences in both underdog win probability and profit margin using multiple statistical tests. Significance was assessed using the Holm-Bonferonni correction \citep{holm} due to the testing of hypotheses for multiple sports.  We used non-parametric tests such as the Kolomogorov-Smirnov (KS) and Mann-Whitney U (MW) test. Non-parametric tests are more appropriate for the underdog profit margin, which exhibits multi-modal behavior and has substantial skew.  

The test results are shown in Table \ref{table:covid_sport}.  The difference in underdog win probability for the NBA is statistically significant for all tests at a 1\% level.  The only other sport showing a significant difference in win probability is the NCAAB, but this is only for the MW test.  The NBA also shows a significant difference in its underdog profit margin for both non-parametric tests.  The NCAAB  underdog profit margin has a significant difference only for the MW test.  However, from Figure \ref{fig:covid_sport} we see that the mean value is negative, which is less interesting from a betting perspective.  

The average NBA underdog profit margin during COVID-19 is the only substantially positive value among all sports.   To verify that this value is truly positive and not just a random fluctuation, we conduct a Wilcoxon signed-rank test \citep{wilcoxon1945} on the NBA COVID-19 games.  This test is a non-parametric version of the standard paired t-test without any assumptions on normality.  The null hypothesis is that the distribution of the values is symmetric about zero.  We find that according to this test, we can reject the null hypothesis at the 1\% level (p-value $\leq 10^{-5}$).  This indicates that the average underdog profit margin is positive for the NBA during COVID-19. We also carried out the test on a season by season basis, and found that the only two seasons for which this result holds was during the 2019-2020 and 2020-2021 COVID-19 seasons. We could not reject the null hypothesis for the  remaining seasons.

\begin{table}
\caption{P-values for different statistical tests for the underdog win probability and profit margin during COVID-19 versus normal time periods segmented by sport ($^*$ indicates significance at a 5\% level and $^{**}$ indicates significance at a 1\% level under the Holm-Bonferonni correction).  The tests are Kolmogorov-Smirnov (KS) and Mann-Whitney U (MW).}
\centering
  \begin{tabular}{|l|c|c||c|c|}
    \toprule
    \multirow{2}{*}{Sport} &
      \multicolumn{2}{|c|}{Underdog win probability} &
      \multicolumn{2}{|c|}{Underdog profit margin} \\
        &KS             & MW                      & KS           & MW          \\\midrule
    NBA & $0.0010^{**}$ & $0.0000^{**}$   & $0.0010^{**}$ & $0.0000^{**}$ \\\hline
    NFL & 1.0000 &0.2937                        & 0.9957& 0.2693 \\\hline
    NHL & 0.5572 &0.0541                       & 0.5572& 0.1895\\\hline
    MLB & 0.8812 &0.1173                       & $0.0001^{**}$& 0.0838\\\hline
    NCAAB & 0.0332 & $0.0006^{**} $     & 0.0207& $ 0.0010^{**} $ \\\hline
    NCAAF & 0.5449 &0.0319                       & 0.5449& 0.0326 \\\hline
    \bottomrule
  \end{tabular}\label{table:covid_sport}
\end{table}

\subsection{Underdog Profit Margin Versus Moneyline Odds for NBA COVID-19 Games}
 We now examine NBA COVID-19 games more closely to understand which games have the highest underdog profit margin.  The underdog moneyline odds cover a wide range.  An interesting question is which segments of this range showed a high underdog profit margin.  To answer this question, we segment the NBA games by their underdog implied win probabilities into bins ranging from zero to one.  Table \ref{table:implied_prob_count} shows the bin intervals and  the number of games in each bin.  
 
\begin{figure}
\begin{center}
\includegraphics[width=\textwidth]{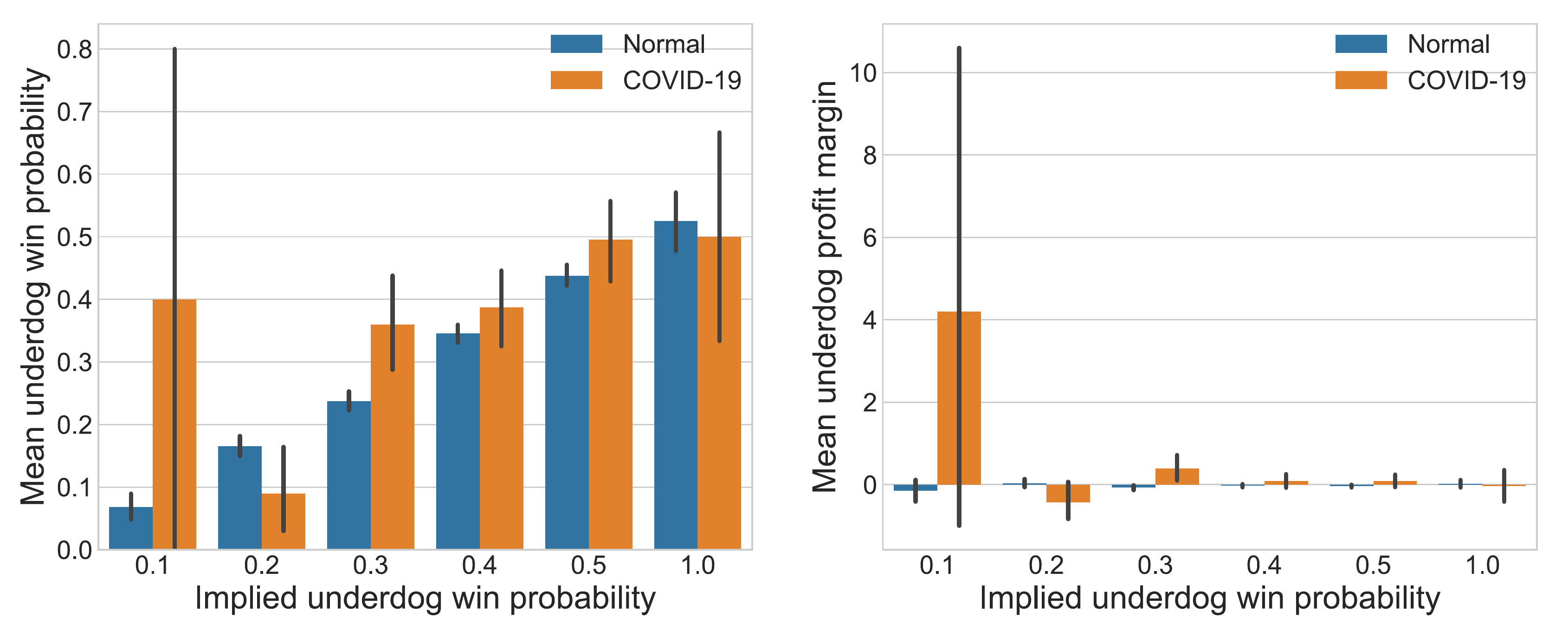}
\caption{Plot of (left) mean underdog win probability and (right) mean underdog profit margin for the NBA grouped by underdog implied win probability (the value on the x-axis is the inclusive upper bound of the implied underdog probability bin.)  The error bars represent 95\% confidence intervals.}
\label{fig:nba_bin}
\end{center}
\end{figure}

 We plot the mean underdog win probability and profit margin versus implied underdog probability bin in Figure \ref{fig:nba_bin}.  We see that both quantities are greater during COVID-19  than during normal seasons for several bins. To asses the statistical significance of these differences, we conducted multiple tests.  The results are shown in Table \ref{table:covid_nba_bins}.  Because we are testing multiple bins simultaneously, we used the Holm-Bonferonni correction to assess significance.  We find that only the $(0.2,0.3]$ bin has a significant difference in both metrics for the MW test at a 1\% level.   

From this analysis we find  evidence that games with implied underdog win probabilities between 0.2 and 0.3 provided much of the inefficiency in the NBA betting markets during COVID-19.  This corresponds to games where the underdog odds are between 233 and 400.  From Table \ref{table:implied_prob_count} we see that there are 153 games in this bin.  This represents 21.7\% of NBA games played during COVID-19.  Therefore, we see that a small fraction of the NBA games played during COVID-19 are responsible for the positive underdog profit margin during this period.

\begin{table}
\caption{Number of NBA games in each implied probability bin for normal and COVID-19 time periods.}\label{table:implied_prob_count}
\centering
  \begin{tabular}{|c|c|c|}
    \toprule
    Implied probability  & Number of NBA games & Number of NBA games \\
     bin &  (Normal) & (COVID-19) \\\hline
    $(0.0,0.1]$ & 539 & 5 \\\hline   
    $(0.1,0.2]$ & 1,955 & 67 \\\hline 
    $(0.2,0.3]$ & 2,603 & 153 \\\hline 
    $(0.3,0.4]$ & 3,627 & 240 \\\hline 
    $(0.4,0.5]$ & 3,445 & 210 \\\hline 
    $(0.5,1.0]$ & 374 & 30 \\\hline 
    \bottomrule
  \end{tabular}
\end{table}

\begin{table}
\caption{P-values for different statistical tests for the NBA underdog win probability and profit margin during COVID-19 versus normal time periods segmented by underdog implied win probability ($^*$ indicates significance at a 5\% level and $^{**}$ indicates significance at a 1\% level under the Holm-Bonferonni correction).  The tests are Kolmogorov-Smirnov (KS) and Mann-Whitney U (MW)}\label{table:covid_nba_bins}
\centering
  \begin{tabular}{|c|c|c||c|c|}
    \toprule
    \multirow{2}{*}{Underdog implied } &
      \multicolumn{2}{|c|}{Underdog win probability} &
      \multicolumn{2}{|c|}{Underdog profit margin} \\
    win probability&KS    & MW              & KS    & MW     \\      \midrule
    $(0.0,0.1]$ & 0.5524 & $0.0025^{*}$           & 0.5524 & 0.0025 \\\hline
    $(0.1,0.2]$ & 0.8359 & 0.0514          & 0.8359 & 0.0537 \\\hline
    $(0.2,0.3]$ & 0.0191 & $0.00021^{**}$  & 0.0186 & $0.0003^{**}$\\\hline
    $(0.3,0.4]$ & 0.8743 &0.1118           & 0.8457 & 0.1139 \\\hline
    $(0.4,0.5]$ & 0.6038 & 0.0652          & 0.5935 & 0.0862 \\\hline
    $(0.5,1.0]$ & 1.0000 & 0.4333          & 0.9999 & 0.2974 \\\hline
    \bottomrule
  \end{tabular}
\end{table}

\subsection{Post-COVID-19 NBA Games}
Though the COVID-19 pandemic continued throughout the NBA 2020-2021 season, there was a point when the league began changing its policies.
The first return from COVID-19 in the 2019-2020 season was played in the isolated bubble.  The beginning of the 2020-2021 season was played in the normal arenas, but with no or very limited fans in attendance.  After the NBA All-Star Game, many teams began allowing larger numbers of fans to attend the games \citep{fans_return}.  We wish to understand the impact of fans on the performance of the betting markets.  

\begin{figure}
\begin{center}
\includegraphics[scale = 0.5]{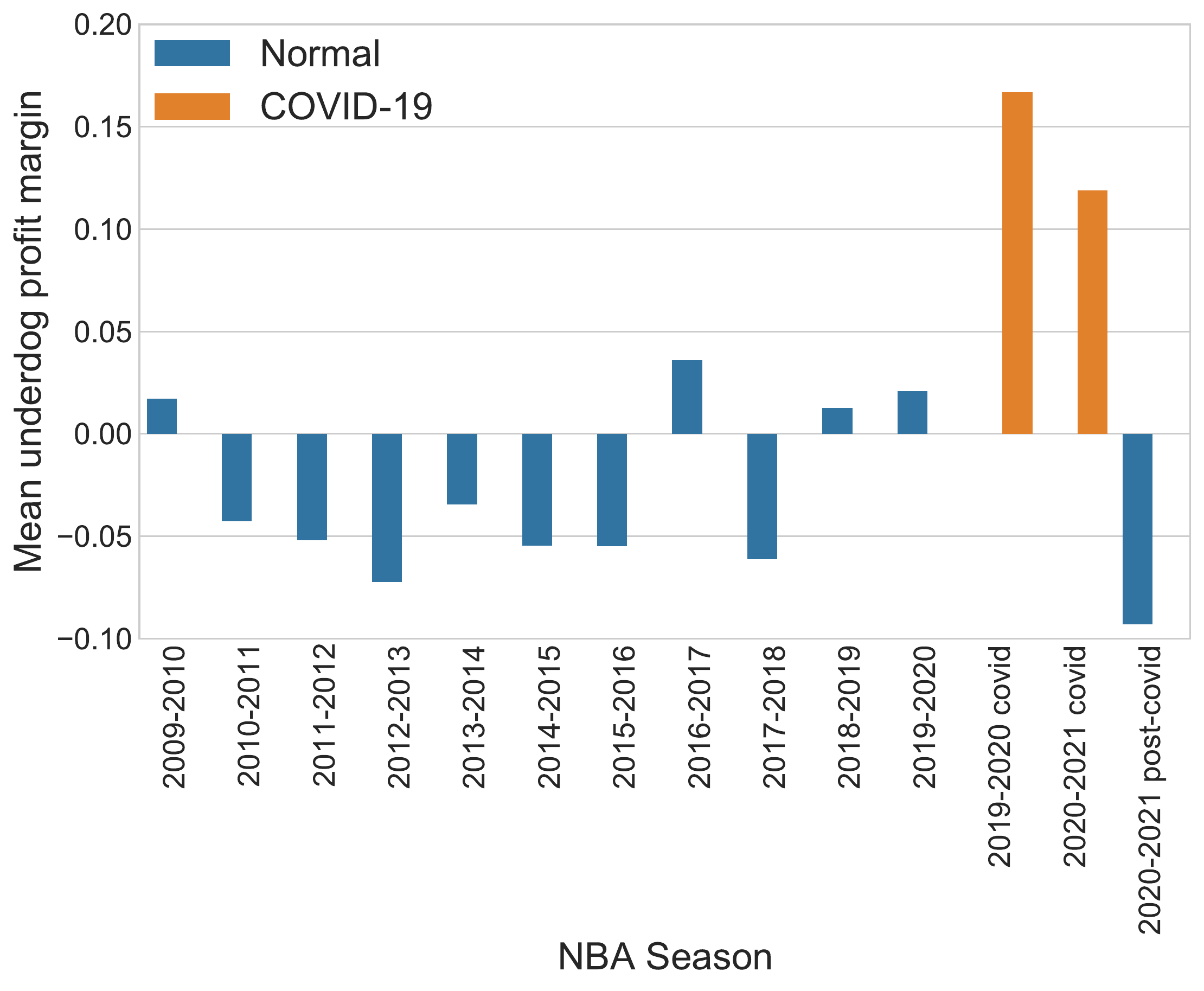}
\caption{Plot of mean underdog profit margin for the NBA versus season and COVID-19 game status. }
\label{fig:NBA_post_covid}
\end{center}
\end{figure}

We plot the average underdog profit margin versus season for the NBA in Figure \ref{fig:NBA_post_covid}.  We see that the average profit margin became very positive in the COVID-19 seasons, but then became  negative during the post-COVID-19 season when fans returned.  We conduct a one-sided Wilcoxon signed-rank test on the two COVID-19 seasons and the post-COVID-19 season using a Bonferonni-Holm correction to verify the sign of these profit margins.  The resulting average underdog profit margin and p-values are shown in Table \ref{table:postcovid}. As can be seen,  the average underdog profit margin is negative for the post-COVID games when fans returned in attendance, while it is positive for the COVID-19 games with no fans (significant at a 1\% level).

\begin{table}
\caption{Average underdog profit margin and one sided p-values for Wilcoxon signed-rank test for the NBA underdog profit margin during the 2019-2020 and 2020-2021 COVID-19 seasons and the 2020-2021 post-COVID-19 seasons ($^*$ indicates significance at a 5\% level and $^{**}$ indicates significance at a 1\% level under the Holm-Bonferonni correction). }
\label{table:postcovid}
\centering
  \begin{tabular}{|c|c|c|}
    \hline
     Season & Average Underdog   \\
            &Profit Margin (p-value)      \\\hline
    2019-2020 COVID-19 & 0.17 ($0.001^{**}$) \\\hline
    2020-2021 COVID-19 & 0.10 ($0.001^{**}$) \\\hline
    2020-2021 Post-COVID-19 & -0.9 ($0.001^{**}$) \\\hline
  \end{tabular}
\end{table}

\subsection{Discussion}
Our analysis has shown that the NBA had large inefficiencies in their moneyline betting markets during COVID-19, while other sports did not.  If we assume that the oddsmakers are good at incorporating known information to set the odds for the games, then any inefficiency must come from some unaccounted factors.  These factors may arise from a combination of the structure of the sports leagues, the  nature of the gameplay, and the impact of COVID-19 on the gameplay. 

We begin with the structure of the leagues.  By structure, we refer to the tendency for underdogs to win games.  In leagues with a concentrated distribution of talent, one would expect underdogs to win more frequently.  In fact, for a league where all teams are equally skilled, we would expect the underdog win probability to be near 0.5.  From Figure \ref{fig:covid_sport}, we see that the sports with the lowest underdog win probabilities are the two college sports (NCAAF and NCAAB).  The spectrum of talent is quite wide in college sports, as they are not professional.  Therefore, one would expect in some games upsets to be incredibly unlikely.  While we did see a small increase in the underdog win probability for the NCAAB, the average underdog profit margin was not positive for both of these college sports.  Therefore, it seems that COVID-19 was not able to impact the games enough to overcome the wide talent spectrum in the college leagues.

We next consider the nature of gameplay.  The highest average underdog win probabilities during normal season games belong to the NHL and MLB, with values exceeding 0.4.  This may be due to how these sports are played.    Table \ref{table:ppg}  shows statistics for the sum of points scored by both teams in a game, which we refer to as \emph{total points}.  We see that the NHL and MLB have the lowest average total points and highest coefficient of variation (standard deviation divided by mean) for total points.  This suggests that scoring is rare and the points scored in a game can fluctuate greatly.  These fluctuations may be due to random factors which cannot be incorporated into the odds.  Baseball involves hitting a small ball thrown at a high velocity with a narrow bat.  Hockey involves hitting a puck into a small net guarded by a goalie.  Randomness plays a large role in both of these sports.  This is likely why the underdog win probability is so large.  We saw that the average underdog profit margin of the MLB and NHL were both negative.  Therefore, it appears that the random factors arising from COVID-19 could not increase the overall randomness such that the betting markets became inefficient.

The NBA's average underdog win probability was near 0.3 during normal seasons, and increased to 0.38 during COVID-19.  From Table \ref{table:ppg} we see that the NBA has the lowest coefficient of variation of total points.  Therefore, NBA basketball is inherently less random than other sports.  The major impact of COVID-19 was to remove live audiences from the games.  In the bubble, there was also no travel and all teams lived in a closed environment.  However, we saw in Figure \ref{fig:NBA_post_covid} that when the NBA went from the bubble to their own arenas, but without fans, the average underdog profit margin stayed positive.  However, once audiences were allowed to attend the games after the All-Star Game, the average underdog profit margin became negative again.  Therefore, it seems likely that the absence of fans at the games was a cause of the betting market inefficiency.  It is not clear why this is the case.  One hypothesis is that when fans are  removed, the home team  advantage is eliminated.   The NBA is a professional league, so it is likely the skill level of the players are concentrated. If this is the case,  then when the home field advantage is absent, the game becomes more susceptible to randomness.     In fact, the average underdog win probability increased during COVID-19 to a value comprable to more random sports such as hockey or baseball.   Our analysis shows that the oddsmakers were not able to account for this, resulting in the inefficiency in the markets.



\begin{table}[t]
\centering
\begin{tabular}{|l|l|r|r|r|r|r|}
\hline
Total Points per  Game & NBA   & NFL  & NHL  & MLB  & NCAAB & NCAAF \\ \hline
Mean         & 205.3 & 45.3 & 5.6  & 8.8  & 139.6 & 55.6  \\ \hline
Standard deviation     & 22.1  & 14.1 & 2.3  & 4.4  & 19.8  & 18.4  \\ \hline
Coefficient of variation   & 0.11  & 0.31 & 0.41 & 0.50 & 0.14  & 0.33  \\ \hline
\end{tabular}
\caption{Statistics of the total points per game (sum of the points scored by each team in a game) for different sports.}\label{table:ppg}
\end{table}

\section{Underdog Betting Strategies}\label{sec:betting}
The analysis in Section \ref{sec:covid} showed that the NBA average underdog profit margin was positive during COVID-19.  This suggests that one could have made profit by betting on NBA underdogs during this period.    From Figure \ref{fig:covid_sport} we see that by just betting an equal amount of money on the underdog in each game results in a 16.7\% profit margin.  In this section we explore how much more profit could be achieved with more complex betting strategies.

\subsection{Betting Scenario}
We consider a scenario where we place daily bets on the underdogs in each game.  The bets for games played in a single day are placed simultaneously.  This is close to what would be done in practice as many games are played at the same time.  The bankroll on day $t$ is denoted $M_t$ and we begin with a bankroll of $M_0$. We use the following strategy to determine the amount of the bankroll to bet each day.  We select a value $\lambda\in[0,1]$, and each day we only reinvest a fraction $\lambda$ of the total bankroll on the games.  

After designating $\lambda M_t$ for betting on day $t$, we must decide how to allocate these funds across the available games.  In practice we would not know a priori which games are profitable.  Therefore, we must consider strategies which only utilize information available before the games begin. For an underdog team $u$ in a game, the information we consider is the implied underdog win probability $p_u$ which is set by the betting market.  We must choose a strategy that maps the $p_u$ to a bet amount.  To do this, we define $w_u\in [0,1]$ as the fraction of the allocated bankroll to bet on the underdog.  With this notation the wager on $u$ is given by $ \lambda M_tw_u$.  To specify $w_u$, we select  a non-negative weight function $f(p_u)$.    Then $w_u$ is given by
\begin{align*}
    w_u = \frac{f(p_u)}{\sum_{v\in G} f(p_v)},
\end{align*}
where  we denote the set of games played on the day as $G$.

There are many possibilities for $f(p_u)$.  If we assume that the implied underdog win probability $p_u$ equals the true win probability $q_u$, then the underdog bet has a mean payout of zero.  In this case, underdog bets can be distinguished by other statistics, such as their variance.  For a 1 USD moneyline bet on an underdog with implied and true win probabilities both equal to $p_u$, the variance is  $(1-p_u)/p_u$.  This means that games with lower underdog win probabilities have a higher variance.  Risk-averse strategies would place smaller wagers on high variance bets, given that the bets have equal means.  This translates to weight functions $f(p_u)$ that are monotonically increasing in $p_u$.  In contrast, risk-seeking strategies would have monotonically decreasing weight functions.     We list several different choices for the weight functions in Table \ref{table:weights}.  We consider both risk-averse and risk-seeking strategies.  The weight functions for risk-seeking strategies are the inverse of the implied underdog win probability, the inverse standard deviation of a Bernoulli random variable for the underdog winning (with $p_u\leq 0.5$), and the standard deviation of a 1 USD underdog moneyline bet.  The risk-averse weight functions are simply the inverse of these functions. We also consider a uniform weight function, which has no risk-preference. 

\begin{table}[h]
\label{table:weights}
\centering
\begin{tabular}{|l|l|}
\hline
\textbf{Name}  & \textbf{Weight Function }  \\ \hline
\textbf{Uniform}       & $1$     \\ \hline
\textbf{Probability}       & ${p_u}$  \\\hline 
\textbf{Inverse Probability}        & $\frac{1}{p_u}$                 \\ \hline
\textbf{Bernoulli }         & $\sqrt{p_u(1-p_u)}$                   \\ \hline
\textbf{Inverse Bernoulli }         & $\frac{1}{\sqrt{p_u(1-p_u)}}$                  \\ \hline
\textbf{Moneyline }       & $\sqrt{\frac{1-p_u}{p_u}}$                 \\ \hline
\textbf{Inverse Moneyline }       & $\sqrt{\frac{p_u}{1-p_u}}$                \\ \hline
\end{tabular}
\caption{Weight functions $f(p_u)$ used in different underdog betting  strategies.}
\end{table}


\subsection{Betting Performance}
\begin{figure}
\begin{center}
\includegraphics[width=\textwidth]{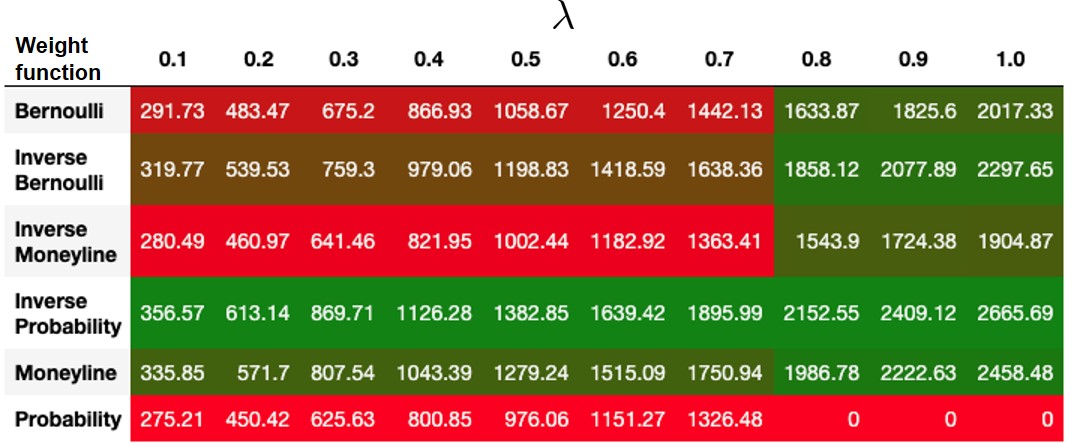}
\caption{Returns [USD] from an initial investment of 100 USD of NBA underdog betting strategies during COVID-19 for different weight functions and values of the reinvestment fraction $\lambda$. A value of zero indicates ruin. }
\label{fig:return_f_lambda}
\end{center}
\end{figure}

\begin{figure}
\begin{center}
\includegraphics[scale = 0.33]{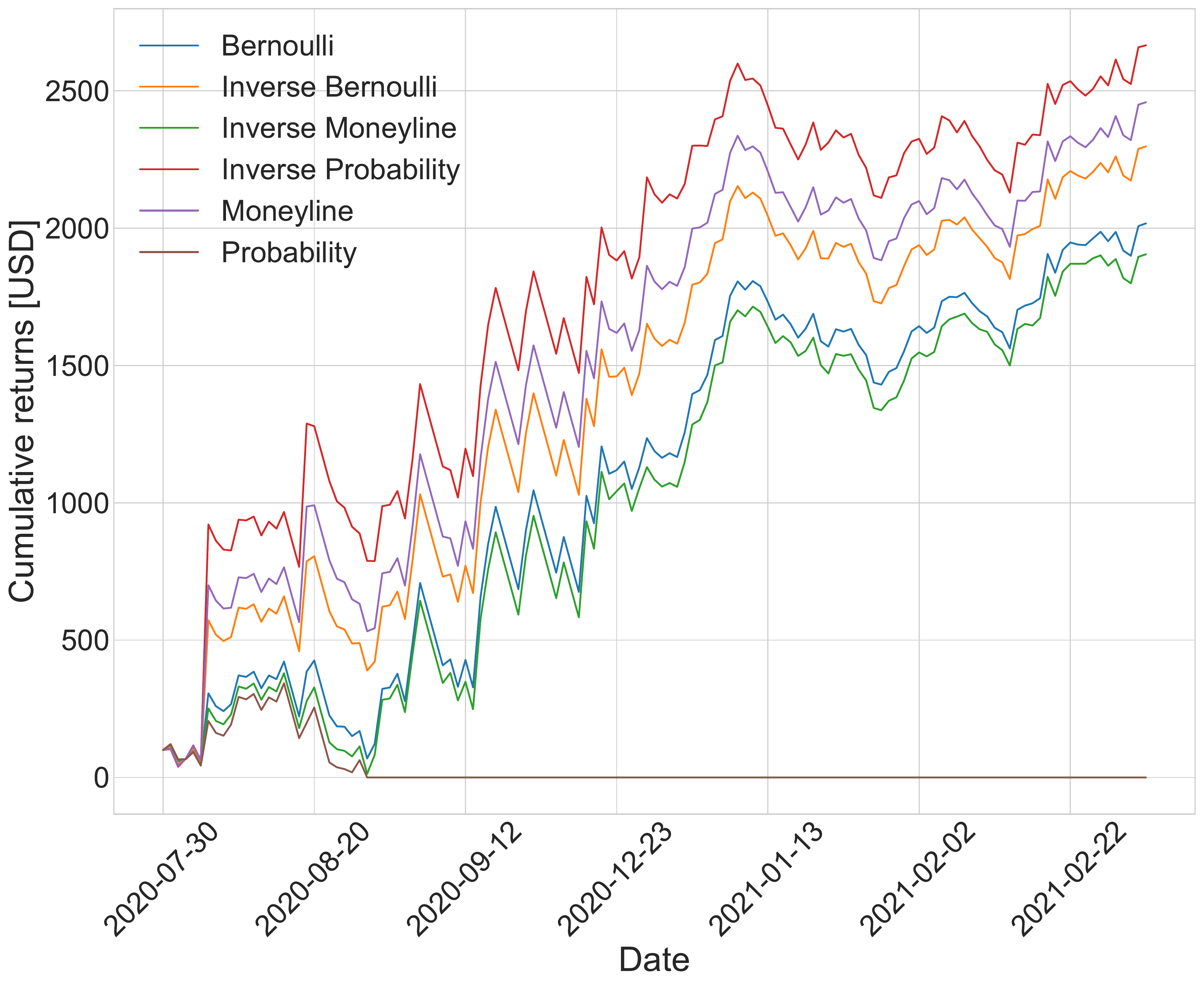}
\caption{Plot of the cumulative  returns of NBA underdog betting strategies  during COVID-19 for different weight functions and a reinvestment fraction of  $\lambda=1.0$. The initial investment is 100 USD.  }
\label{fig:covid_weighting_reinvest}
\end{center}
\end{figure}

We test different betting strategies by varying $\lambda$ and $f(p_u)$.  For $\lambda$ we use values between zero and one spaced 0.1 apart.  We start with a bankroll of $M_0=100$ USD and bets are place on all games in the COVID-19 portion of the 2019-2020 season, and all games in the 2020-2021 season before the All-Star Game.  Figure \ref{fig:return_f_lambda} shows the return for each strategy.  We see that the return is maximized for the strategy where  $\lambda = 1.0$ and the weight function is inverse probability.  The corresponding return is 2,666 USD, or nearly a 26-fold gain in the initial investment. The inverse probability weight is a risk-seeking strategy and places more bets on games with high underdog odds.  However, all weight functions give similar returns when the entire bankroll is bet, except for the probability weight, which ends up losing the entire bankroll.  We compare the time evolution of the returns for each weighting function with $\lambda = 1.0$ in Figure \ref{fig:covid_weighting_reinvest}.  As can be seen, the probability weight function loses the entire bankroll within a month.  All other weight functions show very similar growth for the return over the course of the COVID-19 seasons. 

The final return is not the only metric we can use to evaluate the betting strategies.  Many times one is more concerned with the risk taken to achieve a given return.  One way to quantify this is to use the Sharpe ratio, which equals the mean return divided by the standard deviation of the return.  Strategies with large Sharpe ratios achieve a return with very little risk.  The Sharpe ratio assumes the returns are normally distributed.  However, in the case of moneyline bets, the returns have a bimodal distribution (in fact, for a given game the returns take two values, one negative and one positive).  For this reason, we can define a more robust version of the Sharpe ratio as
\begin{equation}
    \gamma = \frac{\text{median}(R)}{\text{MAD}(R)} \label{eq:gamma}
\end{equation}
where $R$ is the daily return, and MAD is the median absolute deviation. Both the numerator and denominator of this modified Sharpe ration are robust to outliers, making this measure more appropriate for the returns associated with moneyline bets.

\begin{figure}
\begin{center}
\includegraphics[scale = 0.8]{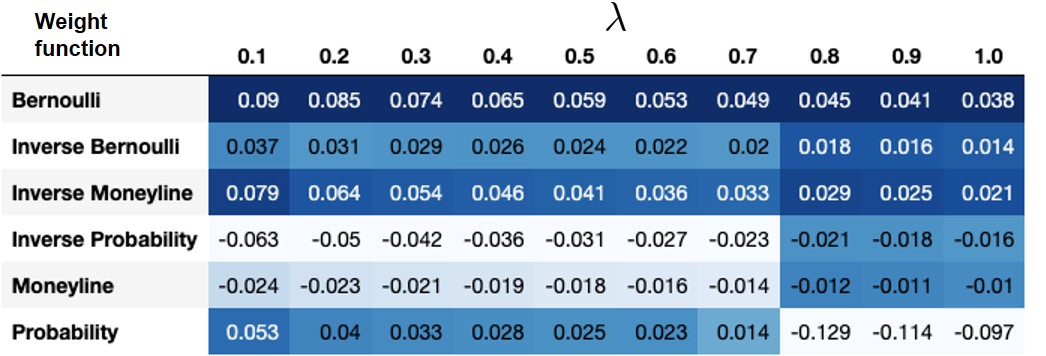}
\caption{Robust Sharpe ratios of NBA underdog betting strategies during COVID-19 for different weight functions and values of the reinvestment fraction $\lambda$.  }
\label{fig:sharpe_robust}

\end{center}
\end{figure}

\begin{figure}[]
\label{fig:normal_sharpe}
\begin{center}
\includegraphics[scale = 0.8]{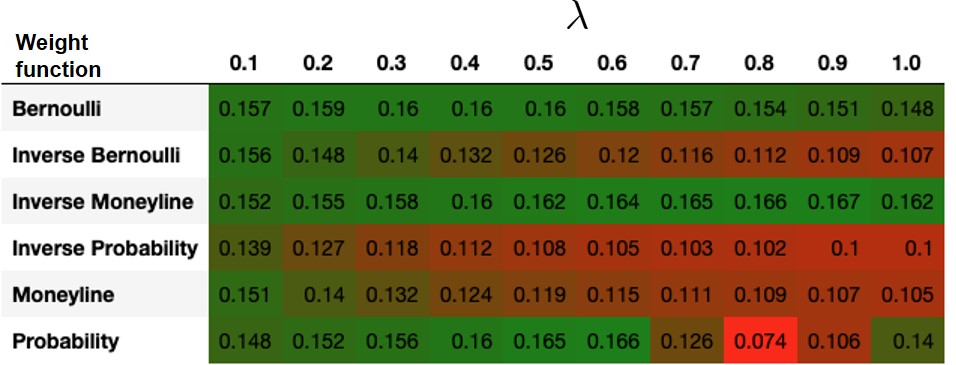}
\caption{Sharpe ratios of NBA underdog betting strategies during COVID-19 for different weight functions and values of the reinvestment fraction $\lambda$.  } 
\label{fig:sharpe}

\end{center}
\end{figure}

We show the robust and normal Sharpe ratios for different betting strategies in Figures \ref{fig:sharpe_robust} and   \ref{fig:sharpe}. We see that $\lambda = 1.0$ does not have the highest Sharpe ratio.  Rather, the highest Sharpe ratios come from the Bernoulli weight with $\lambda$ = 0.1.  This is a risk-averse strategy as most money is not placed on low probability games and only a small fraction of the bankroll is reinvested each day. The return is 291.73 USD for this strategy, which is much lower than the 26-fold gain achieved with a more risk-seeking approach.

\section{Conclusion}\label{sec:conclusion}
COVID-19 affected nearly all professional and college sports.  However, our analysis found that the betting markets for most sports remained efficient, except for the NBA.  Here we saw that there were many more upsets than predicted by the odds makers.  We are not able to precisely identify the reasons for this inefficiency, and it remains an open question as to why it occurred.  However, we do have supporting evidence that it may be due to the more frequent scoring in basketball combined with the absence of fans when the NBA season resumed during COVID-19.  For whatever reason, odds makers were not able to adjust for the impact of these factors.

This NBA market inefficiency provided a lucrative opportunity for bettors.    We found that most of the inefficiency was concentrated around  a small percentage of the games where the underdog had odds between 233 and 400.  Simply betting an equal amount on every underdog resulted in a 16.7\% return, while more complex strategies which combine game specific allocations and reinvestment of winnings resulted in a 26-fold gain in the initial investment. Our work shows that sports betting markets are generally efficient, but occasionally odds makers are not able to correctly account for the impact of extreme events.

\ACKNOWLEDGMENT{The authors thank Professor Toby Moskowitz 
for helpful discussions about this topic. }





\end{document}